\documentclass[aps,preprint,showpacs,amsmath,floatfix,superscriptaddress]{revtex4}

\usepackage{graphicx}

\begin{document}

\title{Exciton Energy Spectra in Two-Dimensional Graphene Derivatives}

\author{Shouting Huang, Yufeng Liang and Li Yang}
%\email[]{lyang@physics.wustl.edu}
\affiliation{Department of Physics, Washington University in St.
Louis, St. Louis, MO 63136, USA}

\date{\today}

\begin{abstract}
The energy spectra and wavefunctions of bound excitons in
important two-dimensional (2D) graphene derivatives, \emph{i.e.},
graphyne and graphane, are found to be strongly modified by
quantum confinement, making them qualitatively different from the
usual Rydberg series. However, their parity and optical selection
rules are preserved. Thus a one-parameter modified hydrogenic
model is applied to quantitatively explain the \emph{ab initio}
exciton spectra, and allows one to extrapolate the electron-hole
binding energy from optical spectroscopies of 2D semiconductors
without costly simulations. Meanwhile, our calculated optical
absorption spectrum and enhanced spin singlet-triplet splitting
project graphyne, an allotrope of graphene, as a candidate for
intriguing energy and biomedical applications.
\end{abstract}

\maketitle

\section{Introduction}

Exciton spectrum, the sequence of electron-hole (\emph{e-h})
binding energies, is the most direct way to understand excitonic
effects of semiconductors. It is also the foundation for
constructing useful models widely used to identify excitonic
effects in optical spectroscopy experiments. For example, the
\emph{e-h} binding energy can be conveniently extrapolated from
the measured sequence of exciton peaks according to model
predictions. In particular, \emph{e-h} interactions are known to
be dramatically enhanced in reduced dimensional structures
\cite{2004Spataru,2005Wang,2006Scholes, 2006Wang,2006Wirtz,
2009yang}. Other than the change of optical spectroscopies, how
these unique quantum confinements influence exciton spectra and
how one subsequently modifies corresponding \emph{e-h} models have
been of fundamental interest. As a result, based on the knowledge
of exciton spectra, numerous exciton models of one-dimensional
(1D) nanostructures \cite{2004Perebeinos,1997Ando, 1994Rinaldi}
and quantum wells \cite{1981Miller,1982Bastard,
1984Greene,1990Andreani,1992Mathieu} have been proposed, which
explain experimental results without costly simulations.

Recently many-electron effects and optical properties of graphene
and its derivatives have ignited substantial interests because of
their unique many-electron effects
\cite{2010Cudazzo,tony2011,2011wang,kats2009}. Because the
thickness of these 2D structures is only a few angstroms, the
perpendicular confinement is extremely strong, making previous
models based on quantum wells (usually with a thickness of tens of
nanometers) inappropriate for these 2D structures. More
importantly, other than studies of the optical absorption, the
exciton spectra of these novel materials are largely unknown.
Therefore, we are unable to extract the general features of
\emph{e-h} interactions and build appropriate exciton models in
these confined 2D systems.

The first-principles simulation based on the many-body
perturbation theory (MBPT) is particularly useful to solve the
above problems because this reliable calculation can provide the
binding energy spectrum of excitons (including dark and bright
states), optical activities, and even their wavefunctions, at the
quantum-mechanical level. This motivates us to employ this method
to calculate excitonic effects in important derivatives of
graphene, \emph{i.e.}, graphyne \cite{1987Baughman,
1998Narita,2011Zhou,2011Kang,2012Srinivasu,2012Malko} and graphane
\cite{2007Sofo,2009Elias,2009Lebegue, 2009Flores,2010Leenaerts}.
First, we expect to reveal the unknown exciton spectra of these
novel 2D structures; secondly, we will build a quantitative model
for identifying excitonic effects of more general 2D
semiconductors without costly \emph{ab initio} simulations,
\emph{e.g.}, extrapolating the \emph{e-h} binding energy, which is
hard to measure directly in experiments.

Beyond fundamental scientific motivations, graphyne, a novel
allotrope of graphene, is of particular interest for optical
applications. Unlike other graphene derivatives, such as graphane
and fluorographene, whose low-energy optical transitions are
depressed by the tetrahedral symmetry \cite{2010Cudazzo}, the
low-energy optical activity of graphyne may be prominent because
of its planar atomistic structure and corresponding active
transitions between $\pi$ electronic states \cite{2011Luo}.
Particularly large-scale graphyne has not been fabricated to date
despite substantial synthesis advances \cite{2011Luo,2000Kehoe,
2001Wan,2003Marsden,2008Haley,2010Diederich,2010Hirsch}. A
quantitative prediction of electric and optical properties of
graphyne is crucial to foresee potential applications and motivate
more research efforts.

In this article, we begin by revealing excited-state properties of
a graphyne structure of current fabrication interest. The
quasiparticle (QP) band gap is appreciable (1.4 eV); the
lowest-energy optical absorption peak is located at 1.0 eV,
meaning a 400-meV \emph{e-h} binding energy; the near-infrared
optical absorbance is more than 6$\%$, making our studied graphyne
one of the most efficient optical absorbers among known materials;
this graphyne structure possesses an impressive spin
singlet-triplet splitting ($\sim$ 150 meV) of excitons. These
features promise exciting energy and biomedical applications.

Moreover, based our calculated exciton spectrum, we propose a
modified one-parameter hydrogenic model, in which the Coulomb
potential is revised to capture the anisotropic quantum
confinement and \emph{e-h} exchange interactions of such a 2D
semiconductor. To justify this model, we have applied it to
graphyne and graphane, achieving excitonic spectra consistent with
\emph{ab initio} results. Therefore, this model may provide a
convenient way to estimate the exciton binding energy without
knowledge of the QP band gap, which shall be of broad interest to
identify many-electron effects from the optical spectroscopy of 2D
nanostructures.

The remainder of this paper is organized as follows: in section
II, we introduce the computing approaches and calculation details;
in section III, quasiparticle band gaps and excitonic effects on
the optical absorption spectrum of graphyne are presented; in
section IV, we present the exciton spectrum of graphyne; in
section V, the modified hydrogenic model is proposed to describe
excitons in 2D semiconductors; in section VI, the proposed model
is applied to explain the exciton spectrum of graphane; in section
VII, we further discuss our exciton model and included
many-electron effects; in section VIII, we summarize our studies
and conclusion.

\section{Computing Setup}

\begin{figure}
\includegraphics*[scale=0.40]{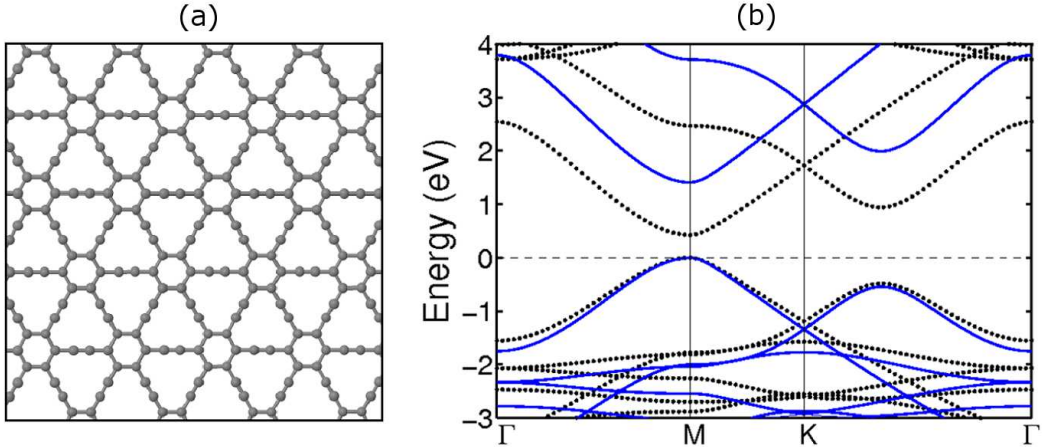}
\caption{(Color online) (a) Top view of the ball-stick model of
our studied graphyne structure. (b) DFT and QP electronic band
structures. The black dots represent the DFT result and the blue
curves are the QP band structure. The top of valence band from
both calculations is always set to be zero.} \label{struc-band}
\end{figure}

The studied graphyne structures is shown in Fig.~\ref{struc-band}
(a), which is predicted by previous studies to be a direct-gap
semiconductor \cite{2012Malko}, a signature for intriguing optical
properties. The ground state is obtained by density functional
theory (DFT)/local density approximation (LDA). The calculations
are done in a plane-wave basis using normconserving
pseudopotentials with a 60 Ry energy cutoff. A coarse 16 x 16 x 1
k-point grid of the first Brillouin zone (BZ) is employed to
compute the self-energy within the single-shot $G_0W_0$
approximation \cite{1986Hybertsen} with a layered Coulomb
truncation. A fine k-grid (64 x 64 x 1) is interpolated from the
coarse grid (16 x 16 x 1) to obtain the converged excitonic states
and optical absorption spectrum by solving the Bethe-Salpeter
Equation (BSE) \cite{2000Rohlfing}. Four valence bands and four
conduction bands are included to calculate optical absorption
spectra of the incident light polarized parallel to the graphyne
plane because of the depolarization effect \cite{2004Spataru,
2009yang}.

\section{Quasiparticle Energy and Optical Excitations of Graphyne}

The DFT and QP band structures are presented in
Fig.~\ref{struc-band} (b), respectively. Because of the depressed
screening in such a 2D semiconductor, enhanced self-energy
correction enlarges the band gap from the DFT predicted 0.43 eV to
1.4 eV, showing an enhanced many-electron correction that is also
observed in other 2D semiconductors \cite{2006Wirtz,2011Luo}. At
the same time, the direct band gap is kept at the M point even
after the GW correction.

\begin{figure}
\includegraphics*[scale=0.25]{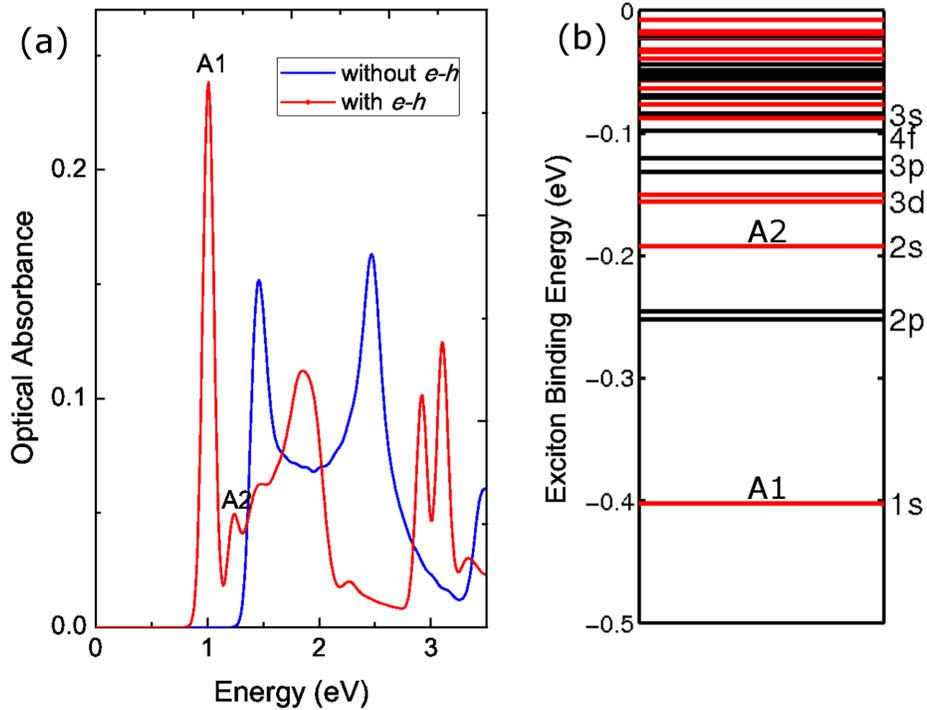}
\caption{(Color online) (a) Optical absorption spectra of graphyne
with and without \emph{e-h} interaction included. The absorbance
value is obtained according to Ref. \cite{2000Rohlfing}. A 0.05 eV Gaussian
broadening is applied to obtain these optical absorption spectra.
(b) Excitonic spectra of bound excitons. The black lines represent
dark states and those red lines represent bright excitons.}
\label{spectra}
\end{figure}

The optical absorption spectra of graphyne are presented in
Fig.~\ref{spectra} (a). In the single-particle absorption spectrum
without \emph{e-h} interactions included (the blue curve), the
optical absorption edge starts from the QP band gap ($\sim$ 1.4
eV) due to the direct-gap nature. More interestingly, a huge
optical absorbance is observed. For example, within the
near-infrared and visible frequency regime, more than 6$\%$ of the
incident light will be absorbed by a single atomic layer, making
our studied graphyne to be one of the most efficient optical
absorbers. This huge optical absorbance is from the significant
overlap between the valence and conduction $\pi$ electronic states
in such a confined structure and consequently enhanced dipole
transitions \cite{2000Rohlfing}.

After including \emph{e-h} interactions, we observe dramatic
excitonic effects on the optical absorption spectrum as shown in
Fig.~\ref{spectra} (a) (the red curve). First, two new absorption
peaks (A1 and A2) appear below the QP band gap because of the
formation of \emph{e-h} pairs (excitons). In particular, the most
prominent exciton with the lowest energy is located at 1.0 eV,
implying a 0.4-eV \emph{e-h} binding energy, which is an order of
magnitude larger than those of excitons in bulk semiconductors.
These enhanced excitonic effects are due to the substantially
depressed screening and quantum confinement
\cite{2004Spataru,2006Scholes,2006Wirtz}.

Moreover, we have calculated the spin-triplet excitons that are
usually dark in the single-photon optical absorption spectrum due
to the selection rule. The lowest-energy spin-triplet exciton is
located at 0.85 eV in the optical spectrum, which is 150 meV below
the first bright singlet exciton (A1) that is located at 1 eV.
Such an enhanced spin singlet-triplet splitting ($\sim$ 150 meV)
is around an order of magnitude larger than those of typical
semiconductors and even carbon nanotubes \cite{2005Perebeinos}.
Since the spin singlet-triplet splitting is decided by the \emph{e-h}
exchange interaction \cite{2000Rohlfing}, the tremendous one observed in
graphyne is from the significant overlap of electron and hole
wavefunctions, which is consistent with the aforementioned huge
optical absorbance.

The above unique optical properties of graphyne may give hope to
numerous potential applications. For example, the strongly bright
exciton A1 located at 1.0 eV \cite{1961Shockley}, the significant
\emph{e-h} binding energy ($\sim$ 400 meV) and impressive spin
singlet-triplet splitting give hope to potential PV materials
\cite{2007Peet,2009Park}. Other than energy applications, our
studied graphyne structure exhibits an extremely strong absorbance
between 1 eV and 2 eV, which may be of interest for biomedical
applications \cite{2011hessel}. It has to be pointed out that we
only predict the fundamental properties of graphyne. For realistic
applications, many other factors, such as electron-phonon
interactions, defects and mobility, will be crucial.

\begin{figure}
\includegraphics*[scale=0.50]{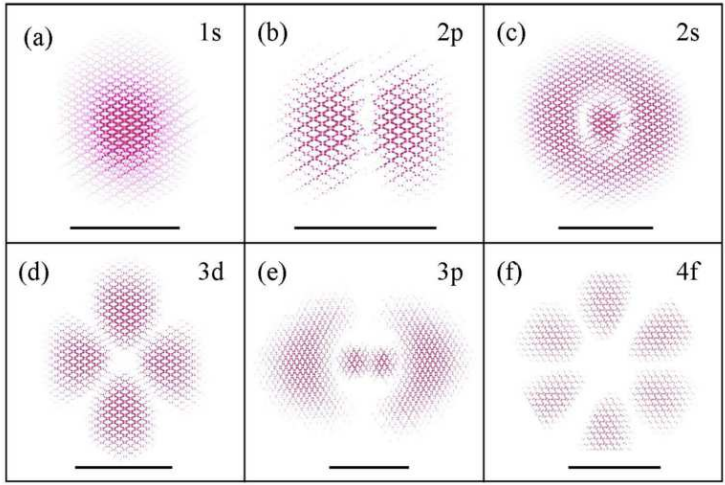}
\caption{(Color online) Top views of the square of the electron
wavefunctions of the characteristic bound excitons of graphyne
from Fig. 2 (b). The hole is fixed at the center of each plot. The
real-space 10-nm scale bars are presented, respectively.}
\label{exciton-1}
\end{figure}

\section{Exciton Spectrum of Graphyne}

Beyond focusing on these optically prominent excitons, it is
necessary to study the whole exciton spectrum, which is crucial to
understand \emph{e-h} interactions and subsequent modelling
efforts. In Fig.~\ref{spectra} (b) we lists all bound exciton
states of graphyne according to their binding energy. Because the
direct band gap is located at M points of the first BZ, each
exciton energy level is actually triple-degenerated, but here we
only consider one set of them. An immediate question is raised
from Fig.~\ref{spectra} (b); the second lowest energy levels are
doubly degenerated and dark in the optical absorption spectrum.
This substantially conflicts the usual hydrogenic exciton model,
in which the lowest two excitons shall be the non-degenerated and
bright 1$s$ and 2$s$ states, respectively.

In order to understand this unusual exciton spectrum, we first
focus on their real-space wavefunctions. In Fig.~\ref{exciton-1}
the six lowest-energy excitonic states are plotted (for
degenerated states, we only plot one of them). We see the
distributions of wavefunctions are similar to the hydrogenic
model, \emph{e.g.}, the spherical symmetry of the $s$ orbital,
those angular momentum characters of $p$, $d$ and $f$ orbitals,
and their nodal structures. As a result, we identify these states
with the same parities as the hydrogenic model, \emph{i.e.}, 1$s$,
2$s$, and 2$p$, \emph{etc.}, as marked in Fig.~\ref{spectra} (b).
The optical selection rules on these states are also almost
preserved. For example the $s$ states are bright while the $p$
states are dark. The only exception is the 3$d$ states, which
shall be dark while they are slightly bright in Fig.~\ref{spectra}
(b). This is due to the fact that the calculated graphyne
structure is only quasi-2D, which cannot keep the perfect
symmetry.

On the other hand, the order of these exciton states in
Fig.~\ref{spectra} (b) is 1$s$, 2$p$, 2$s$, 3$d$, 3$p$ and 4$f$,
\emph{etc.}, which is qualitatively different from exciton spectra
of either 2D or 3D hydrogenic model. Moreover, if we fit the
energy dependence of those bright $s$ states according to the main
quantum number $n$, the first-principles result decays much more
slowly than the $\frac {1} {(n-0.5)^2}$ relation of 2D hydrogenic
model or the $\frac {1} {n^2}$ relation of 3D hydrogenic model.
These similarities and dissimilarities between \emph{ab initio}
results and hydrogenic models encourage us to modify the
hydrogenic model by approximating the perpendicular confinement.

\begin{figure}
\includegraphics*[scale=0.45]{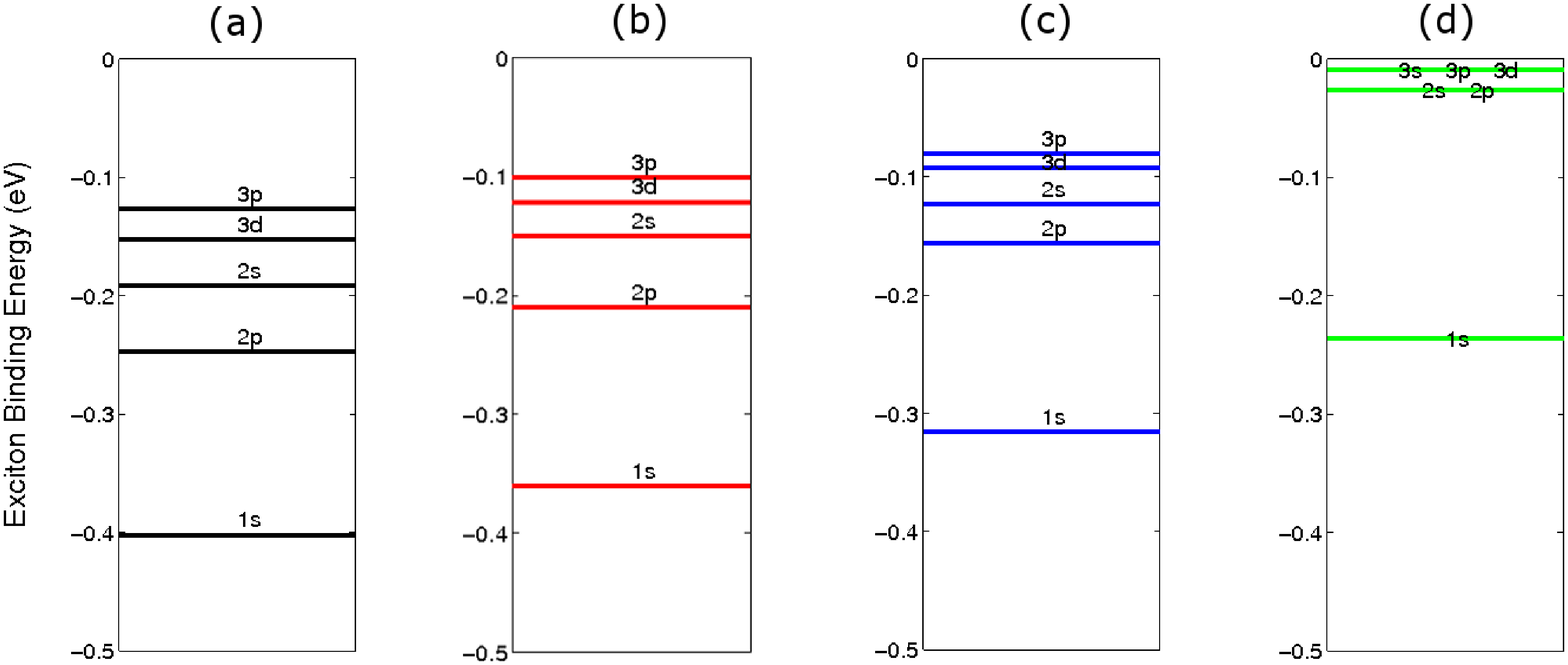}
\caption{(Color online) Exciton spectra. (a), (b), (c), and (d)
are results of graphyne from the \emph{ab initio} simulation, our
model ($m^*=0.071 m_0, d_0=2.44 nm$), the model from
Ref.~\cite{2011rubio}, and the original 2D hydrogenic model,
respective.} \label{compare-1}
\end{figure}

\section{An Exciton Model in 2D Semiconductors}

An obvious improvement to the typical 3D hydrogenic model is to
confine the Coulomb interaction within a finite width
perpendicular to the graphyne layer. In particular, the typical
size of excitons shown in Fig.~\ref{exciton-1} is around 10 nm,
which is much larger than the thickness of the electron
distribution perpendicular to the graphyne plane ($\sim$ a few
\AA). This validates the first-order approximation that the
thickness of graphyne is a small number compared to the average
distance between electron and hole. As a result, we introduce the
following modified Coulomb interaction:
\begin{equation}
V(r) = - \frac {1}{\varepsilon_0} \frac {1} {\sqrt{r^2 + d_0^2}},
\label{BSE-1}
\end{equation}
where $r$ is the polar radius of cylindrical coordinates and $d_0$
is the parameter to reflect the effective thickness of 2D
excitons. Actually this type of Coulomb interaction had been
applied to study many-electron systems before
\cite{2002Lozovik,2005tan,2008Schindler}. With the help of the
separation of variables, all exciton levels can be obtained by
solving a 1D single-particle Schrodinger equation
(Eq.~(\ref{BSE-2}), in Hartree atomic units) by the finite-element
simulation,

\begin{equation}
[\frac {1}{2m^*}(- \frac {d^2}{dr^2} - \frac {1}{r} \frac{d}{dr} +
\frac{l^2}{r^2}) - \frac {1}{\sqrt{r^2 + d_0^2}}]\cdot R(r)= E
\cdot R(r).
\label{BSE-2}
\end{equation}

The effective mass $m^*$ is the reduced mass of electrons and
holes (averaged by all in-plane directions), which can be obtained
by simple DFT calculations because many-electron corrections
usually do not change the curvature of electronic bands
significantly. In a word, only one parameter, the effective
thickness $d_0$, is essential in this model.

In realistic cases, we optimize $d_0$ according to the energy
spacing between the first two bright singlet excitonic (1$s$ and
2$s$) states, which shall be the easiest data from the optical
absorption or luminescence spectrum experiments. In this work, as
a example, we fit $d_0$ according to the energy spacing of 1$s$
and 2$s$ states from the \emph{ab initio} simulated optical
absorption spectrum shown in Fig.~\ref{spectra} (a) ($A_1$ and
$A_2$). The results are concluded in Figs.~\ref{compare-1}, in
which this modified hydrogenic model provides surprisingly good
explanations. The deviation of the binding energy between the
model and \emph{ab initio} result is less than 40 meV. Considering
the extremely light simulation of the model, this model shall be
of help for researchers who are not experts of the
first-principles MBPT. Besides the binding energies, the
eigenstates of the model exhibit exactly the same energy order as
the results from \emph{ab initio} simulation.

\begin{figure}
\includegraphics*[scale=0.50]{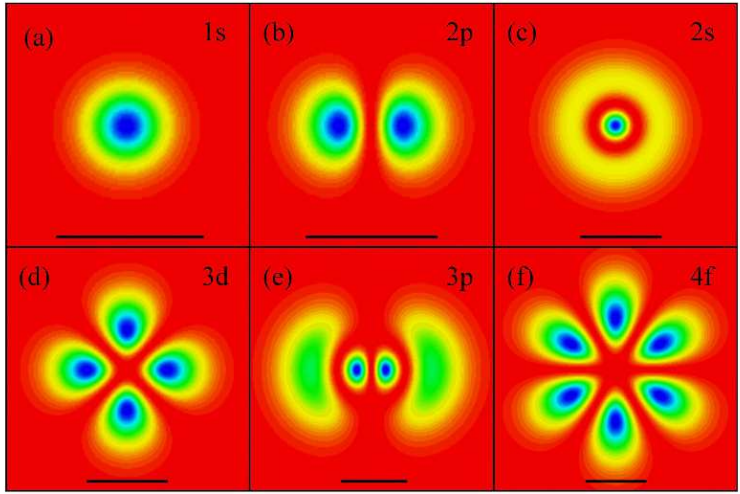}
\caption{(Color online) Top views of the square of the electron
wavefunctions of the characteristic bound excitons of graphyne
solved by our model. The hole is fixed at the center of each plot.
The real-space 10-nm scale bars are presented, respectively.}
\label{wave-2}
\end{figure}

In Fig.~\ref{wave-2}, we have presented the wavefunctions of those
excitonic states solved from Eq.~\ref{BSE-2}, using the parameter
of graphyne. More surprisingly, this model even gives the
similarly sized wavefunctions of these excitons ($\sim$ 10nm)
compared to those first-principles results, in addition to the
same nodal structures.

We have compared our results with another recently proposed model
for describing \emph{e-h} interactions with negligible exchange
interactions in 2D semiconductors \cite{2011rubio, 1979jetp} in
Fig.~\ref{compare-1} (c). This model also gives reasonably good
predictions; the \emph{e-h} binding energy is around 320 meV,
around 80 meV less than the \emph{ab initio} result. As shown in
Fig.~\ref{compare-1}, our model provides better results. This is
not surprising because our model has a fitted parameter $d_0$
while the model from Ref.~\cite{2011rubio} does not have tunable
parameters. In particular, $d_0$ is fitted from singlet states and
it thus more aptly includes subtle many-electron effects, such as
\emph{e-h} exchange interactions. This brings new physical
meanings to the parameter $d_0$ in addition to the thickness
effect. Further discussion will be presented in Section VII.

\section{Application of the exciton model to Graphane}

\begin{figure}
\includegraphics*[scale=0.45]{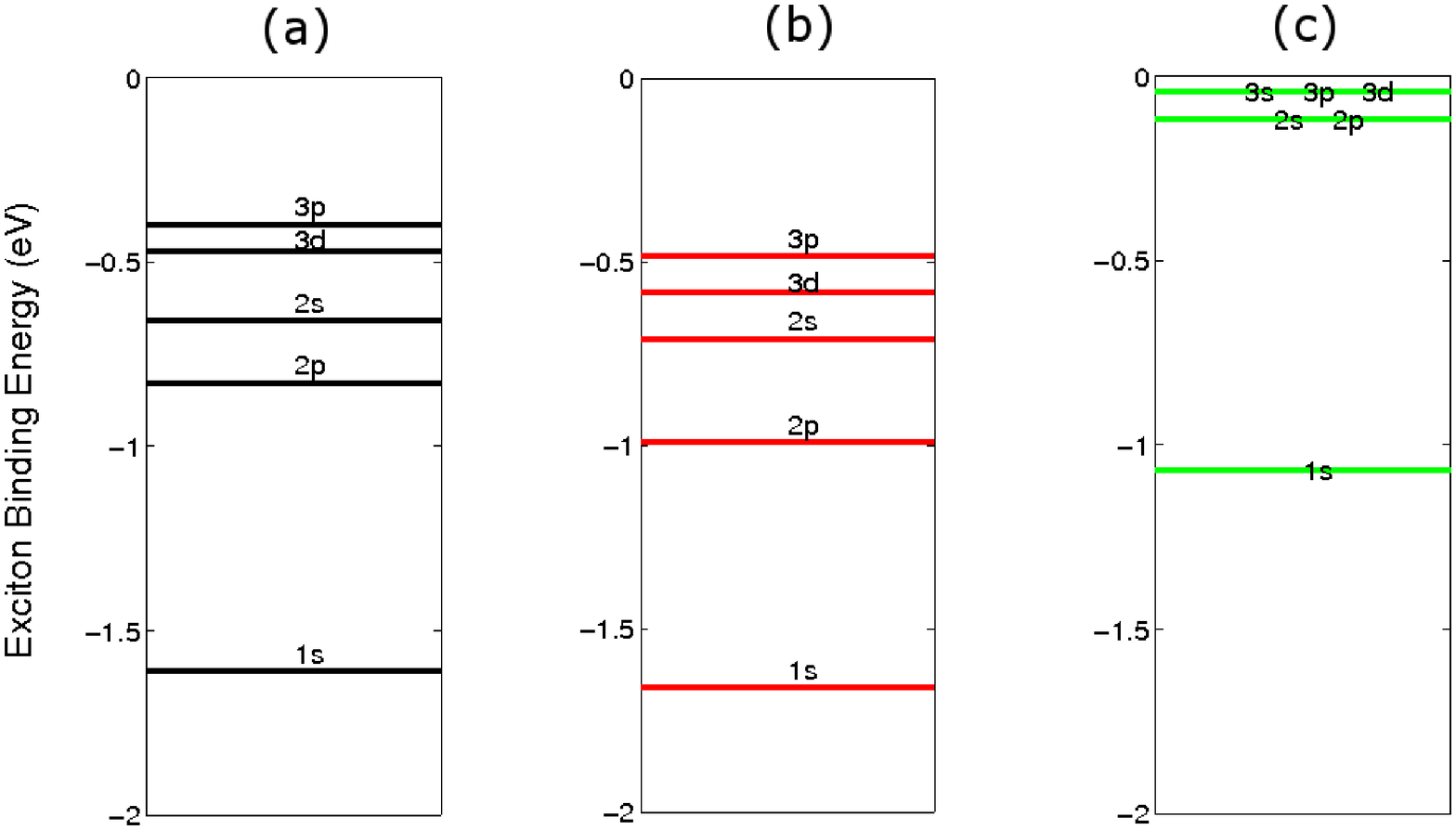}
\caption{(Color online) Exciton spectra. (a),(b) and (c) are
results of graphane from the \emph{ab initio} simulation, our
model ($m^*=0.353 m_0, d_0=0.54 nm$), and the original 2D
hydrogenic model, respective.} \label{compare-2}
\end{figure}

Meanwhile, we also calculate excitonic spectra of another
important 2D graphene derivatives, hydrogen-passivated graphene
(graphane). Here the lowest-energy exciton of graphane is a
charge-transfer and relatively dark one \cite{2010Cudazzo}, which
is qualitatively different from the bright and non-charge-transfer
exciton in graphyne. This difference provides us a good
opportunity to justify the application range of our model. The
results are concluded in Figs.~\ref{compare-2}. We again obtain
the excellent consistence between exciton spectra from both the
model and \emph{ab initio} simulation. For the comparison purpose,
we have listed the results from the original 2D hydrogenic model,
which exhibit substantially larger errors for both graphyne and
graphane: the binding energy is much smaller than that from
\emph{ab initio} simulations; the degeneracy of excitons is not
correct and the order of exciton energy levels are qualitatively
wrong.

We have compared our results with the previous model
\cite{2011rubio} as well. Because of the lack of exchange
interactions due to the charge-transfer nature of involved
excitons, all models give the similar result about the binding
energy of excitons in graphane, around 160 meV, which agrees well
with the \emph{ab initio} result.

\section{Further Discussion of the Exciton Model}

\begin{figure}
\includegraphics*[scale=0.40]{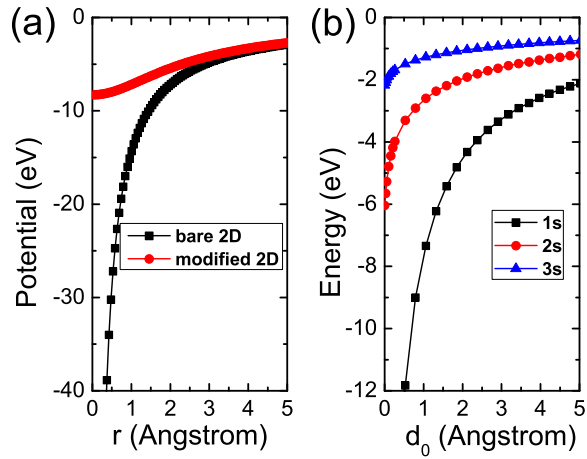}
\caption{(Color online) (a) Bare Coulomb potential and our
modified potential for \emph{e-h} interactions.(b) The evolution
of the binding energy of 1$s$, 2$s$, and 3$s$ states according to
the effective thickness $d_0$ from the solution of
Eq.~(\ref{BSE-2}) when $m^*=m_0$.} \label{potential}
\end{figure}

Our model can provide more of systematic knowledge of excitons in
2D semiconductors. We have plotted the potential profiles of bare
Coulomb potential and our modified \emph{e-h} interaction
potential in Fig.~\ref{potential} (a). They are significantly
different from each other when $r$ is small, e.g., $r$ is less
than 2 \AA. This hints us the most significant corrections from
our model is for those smaller-sized exciton states. Furthermore,
we have present how the binding energy of the first three $s$
exciton states evolves with effective thickness ($d_0$) from the
solution of Eq.~(\ref{BSE-2}), where the effective mass $m^*=m_0$
(for other $m^*$ values, the binding energy and effective
thickness can be scaled by $m^*$, respectively), in
Fig.~\ref{potential} (b). This shows the quantum confinement
effects on \emph{e-h} pairs. For example, we can see the energy
spacings between these $s$ states shrink as we increase $d_0$.
This explains why graphyne and graphane have a slower decaying
trend of the exciton binding energy than that of original 2D
hydrogenic model. Finally, it has to be pointed out that our model
only work well with 2D semiconductors with the direct band gap,
whose effective masses of electrons and holes are not extremely
anisotropic.

It is prudent to identify additional physical mechanisms that are
tied to the fitted parameter $d_0$, which was originally
introduced by the finite thickness of these 2D semiconductors. We
find that $d_0$ is around 2.44 nm in graphyne, which is too big to
be regarded as the effective thickness. However, because $d_0$ is
fitted by the energy spacing of the two lowest-energy singlet
excitons, it must be influenced by \emph{e-h} exchange
interactions. In particular, exchange interactions usually serve
as a repelling force (this is why the energy of spin triplet
states is usually lower than that of singlet states), which will
reduce the \emph{e-h} attraction and, in our model, enlarge the
parameter $d_0$ consequently. In this sense, a larger $d_0$ also
reflects enhanced exchange interactions through the fitting
process. This is consistent with the large spin singlet-triplet
splitting in graphyne. For graphyne, the small $d_0$ is around
0.54 nm, which corresponds to its negligible exchange interaction.

This helps us clearly understand the solutions from
Eq.~\ref{BSE-2}, which may provide either spin-singlet results or
spin-triplet results, depending on how one obtains the fitted
parameter $d_0$. In our study, $d_0$ is obtained by fitting
singlet states. Thus the solutions must be interpreted in the
context of singlet states because the exchange interaction is
implicitly included through the fitting process.

\section{Conclusions}

In conclusion, we perform the first-principle GW-BSE approach to
study optical excitations of graphyne. Our calculation reveals
that graphyne is a promising material which may own the potential
for a wide range of applications, e.g., PV and photo therapy.
These quantitative prediction shall be of importance to spur more
research resource and interest to graphyne. At the same time, we
analyze the excitonic spectra of graphyne and propose a modified
hydrogenic model that not only explains the exciton spectrum of
graphyne but also that of graphane, shedding light on a convenient
approach to understanding excitonic spectra and estimate the
binding energy of excitons in 2D semiconductor.

\section{ACKNOWLEDGMENTS}

This research is supported by NSF Grant No. DMR-1207141. The
computational resources have been provided by Lonestar of Teragrid
at the Texas Advanced Computing Center. The ground state
calculation is performed by the Quantum Espresso
\cite{2009Giannozzi}. The GW-BSE calculation is done with the
BerkeleyGW package \cite{2012Deslippe}.

\end{document}